\documentclass[11pt]{article}
\usepackage{moriond,epsfig,wrapfig}

\def\be{\begin{equation}}
\def\ee{\end{equation}}
\def\bea{\begin{eqnarray}}
\def\eea{\end{eqnarray}}
\def\lsim{\mbox{\raisebox{-.6ex}{~$\stackrel{<}{\sim}$~}}}

\begin{document}
\rightline{CERN-PH-TH/2010-101}
\vspace*{4cm}
\title{EXCITED DARK MATTER VERSUS PAMELA/FERMI}

\author{J.M.\ CLINE}

\address{CERN Theory Division, CERN, 
Case C01600, CH-1211 Gen\`eve, Switzerland\\
and\\ McGill University, Department of Physics, 3600 University
St., Montr\'eal, Qc H3A2T8, Canada}

\maketitle\abstracts{
Excitation of multicomponent dark matter in the
galactic center has been proposed as the source of low-energy
positrons that produce the excess 511 keV $\gamma$ rays that have been
observed by INTEGRAL.  Such models have also been promoted to explain
excess high-energy $e^\pm$ observed by the PAMELA, Fermi/LAT and 
H.E.S.S.\ experiments.  We investigate whether one model can simultaneously fit
all three anomalies, in addition to further constraints from inverse
Compton scattering by the high-energy leptons.  We find models that
fit both the 511 keV and PAMELA excesses at dark matter masses
$M < 400$ GeV, but not the Fermi lepton excess.  The conflict arises
because a more cuspy DM halo profile is needed to match the observed
511 keV signal than is compatible with inverse Compton constraints
at larger DM masses.
}

\section{Galactic cosmic ray anomalies and DM collisions}

There are several hints of unexplained sources of electrons and
positrons in our galaxy, which could be due to collisions of dark
matter (DM).  The longest-standing one is the excess of 511 keV
$\gamma$ rays from the galactic center, first seen by balloon-borne
detectors in the 1970's, and most recently measured by the SPI
spectrometer aboard the INTEGRAL satellite (for a review, see
ref.\ \cite{diehl}).  More recently, a number of experiments have found
evidence for $e^\pm$ at higher energies, in excess of those
understood to be coming from known sources.  Among these, 
PAMELA\cite{pamela} reports an excess in the positron fraction at energies
of $10-100$ GeV, while the Fermi Large Area Telescope 
(LAT)\cite{fermi}  and H.E.S.S.\cite{hess} observe an excess of
$e^++e^-$ in the $100-1000$ GeV energy range.  

Although many different astrophysical explanations have been proposed
as the source of the low-energy positrons that produce the 511 keV
signal, there is no consensus.\footnote{for example, the argument of 
ref.\ \cite{lmxrb} that low-mass x-ray binaries are most likely source has
been criticized in ref.\ \cite{silk}.}  Pulsars have been
proposed as a likely source of the PAMELA and Fermi leptons (see for
example refs.\ \cite{hooper,grasso}) but the uncertainties in the
parameters characterizing such sources still leave room  for other
interpretations.

\begin{wrapfigure}{r}{0.25\textwidth} 
\psfig{figure=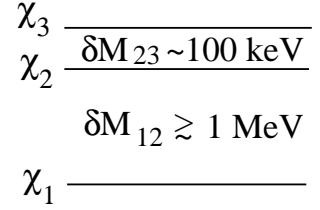,width=0.25\textwidth}
\caption{ Inverted mass hierarchy for excited DM states.}
 \label{spect}
\end{wrapfigure}

Although DM annihilations had previously been suggested as the source
of some of these anomalies, ref.\ \cite{AH} was the first to point
out a class of DM models that could potentially explain all of them
(and a few others: the WMAP haze and the DAMA/LIBRA annual
modulation). Namely, these are models where the DM has a mass $M$
near the TeV scale, and has several components that acquire naturally
small mass splittings $\delta M\lsim$1 MeV from radiative corrections.
A new hidden sector Higgs or gauge boson with mass $\mu\lsim$1 GeV
mediates annihilations of the DM into $e^\pm$ but not antiprotons
(since $\mu<2m_p$), of which no excess has been observed by PAMELA. 
All of this can be economically achieved by assuming the hidden
sector gauge symmetry is nonabelian and spontaneously breaks near the
GeV scale. Then the mediator is one of the gauge bosons $B_{\mu}$,
which can mix with the standard model hypercharge $Y_\mu$ through the
dimension-5  gauge kinetic mixing operator  $\Lambda^{-1}\Delta^a
B^a_{\mu\nu} Y^{\mu\nu}$, where $\Delta^a$ is a hidden sector Higgs
field in the adjoint representation that gets a VEV.  Some of the 
simplest examples involving SU(2) gauge symmetry were considered by us
in ref.\ \cite{twist,nonabelian}.

\section{Exciting Dark Matter in the Galactic Center}

The excited DM mechanism (XDM) for explaining the 511 keV excess was first
proposed in ref.\ \cite{FW}.  The ground state DM particles undergo
inelastic scattering to the excited state by
$\chi_1\chi_1\to\chi_2\chi_2$, followed by decays $\chi_2\to\chi_1
e^+ e^-$ into nonrelativistic $e^\pm$.   However a quantitative
computation of the excitation cross section was not used there, and
ref.\ \cite{PR} argued that the rate of excitation was too small to 
account for the observations unless many partial waves were at their
maximum values allowed by unitarity. 

In refs.\  \cite{twist,ccffr} we numerically computed the
excitation cross section by solving the Schr\"odinger equation, and
showed that indeed the suspicion of ref.\ \cite{PR} was correct, the rate
of $e^+$ production is too small, even varying all the model
parameters and DM halo properties over a wide range.

\begin{figure}[h]
\centerline{\psfig{figure=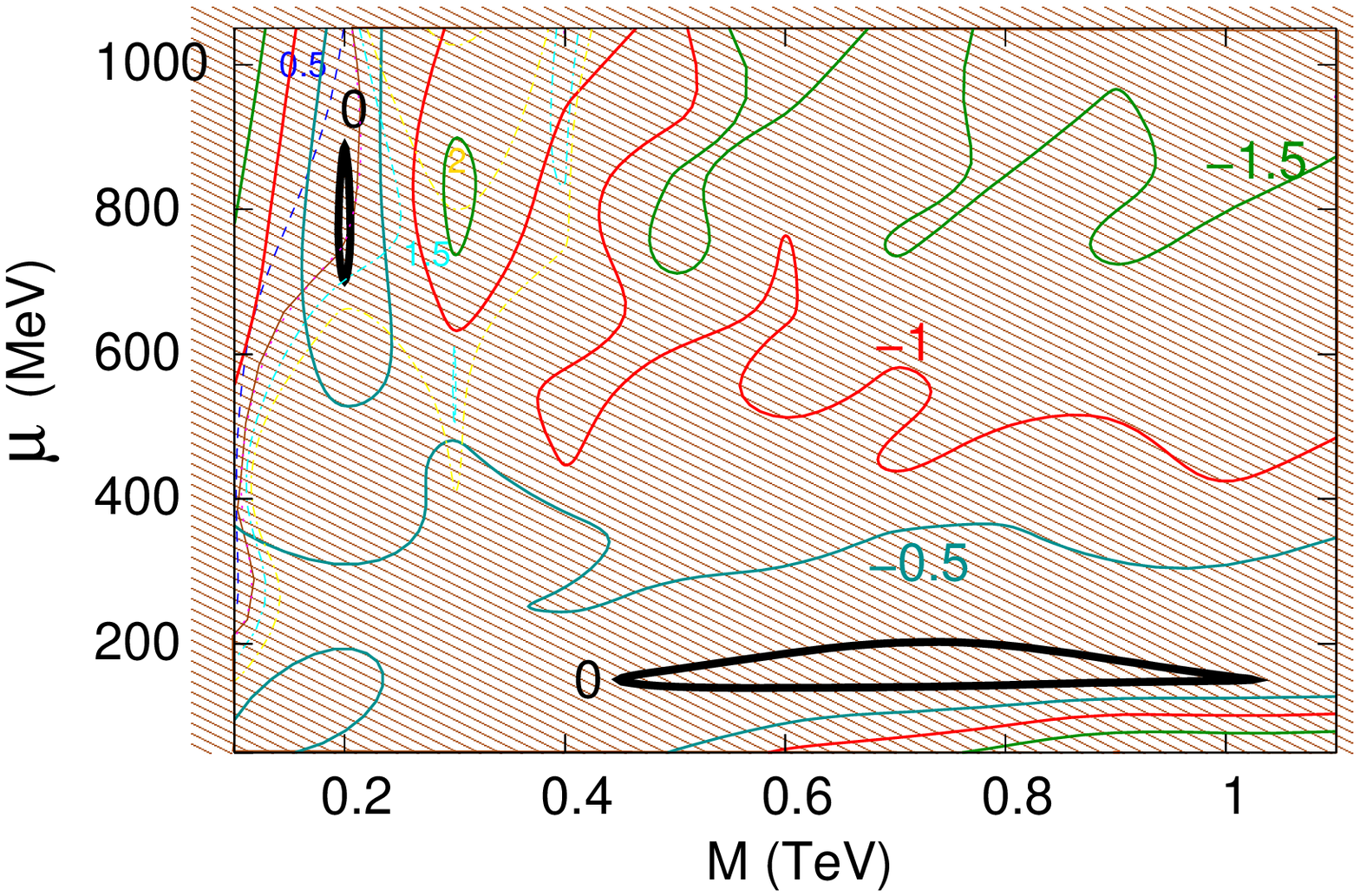,width=0.51\textwidth}
\hfil
\psfig{figure=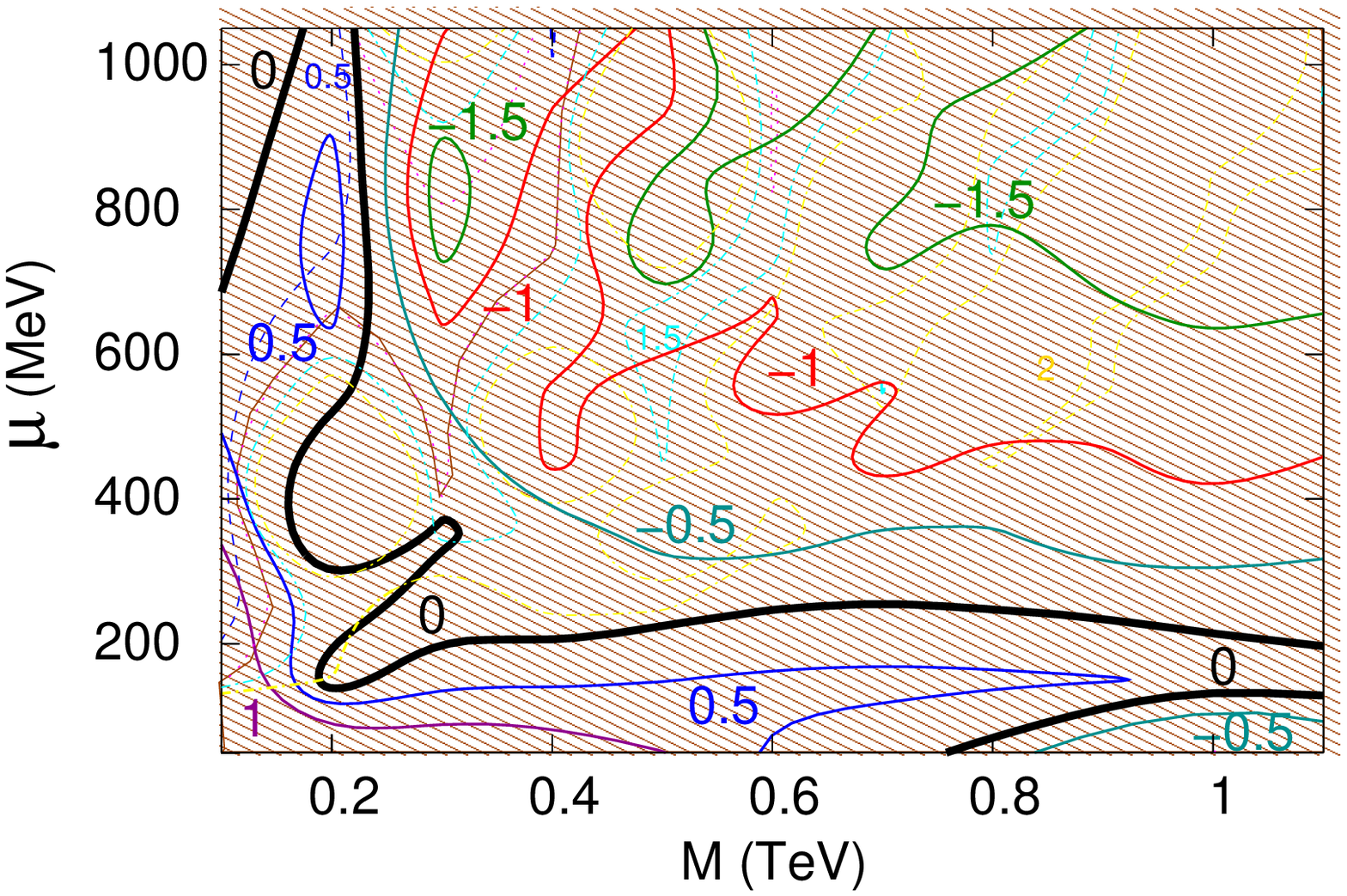,width=0.5\textwidth}} 
\caption{Left: contours of the
rate of positron production, $\log(R_{e^+}/ R_{\rm obs})$
(for INTEGRAL 511 keV $\gamma$ rays) in plane of
gauge boson mass $\mu$ versus DM mass $M$ for $\delta M_{23}=100$ keV
mass splitting and halo Einasto profile parameters $\alpha=0.17$, $r_s=15$
kpc, $\rho_\odot=0.4$ GeV/cm$^3$, $v_0=250$ km/s.$^{18}$\   Heavy 
contours match the observed rate.  Dashed curves are contours of 
inverse Compton (IC) signal over IC bound.  Shaded regions
are excluded by IC constraint.  Right: same, but with
$\delta M_{23}=25$ keV, $\alpha=0.20$, $r_s=15$
kpc, $\rho_\odot=0.3$ GeV/cm$^3$, $v_0=220$ km/s.}
 \label{xdmres}
\end{figure}

\noindent At the same
time, we proposed a solution, involving the existence of a stable
excited state that undergoes scattering 
$\chi_2\chi_2\to\chi_3\chi_3$, followed by the decay $\chi_3\to\chi_1
e^+ e^-$.  This can have a smaller mass gap $\delta M_{23}\sim 100$
keV which is easier to excite in DM collisions than the larger one
$\delta M_{13}>2 m_e$.   This ``inverted mass hierarchy'' is 
shown in figure \ref{spect}.

\begin{wrapfigure}{r}{0.4\textwidth} 
\psfig{figure=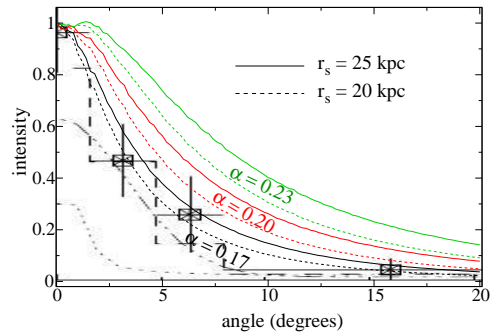,width=0.4\textwidth}
\caption{Observed angular distribution of INTEGRAL 511 keV signal,
and theoretical predictions for different Einasto parameter
values $\alpha=0.17,0.20,0.23$, $r_s=20,25$ kpc.
\centerline{\ }
\centerline{\ } }
 \label{angle}
\end{wrapfigure}

Figure \ref{xdmres} (left panel) shows an example of our new 
contours for the rate of
positron production compared to the observed rate in the $M$-$\mu$
plane,\cite{CCF}\  using the mass splitting $\delta M_{23}= 100$
keV, and the gauge coupling  $\alpha_g = 0.031\, (M/{\rm TeV})$ required
for getting the right relic density.\cite{nonabelian}\ \  The DM 
density profile is taken to be of the Einasto form,
\be
	\rho = \rho_\odot \exp\left[ -{2\over\alpha}\left(
	\left(r\over r_s\right)^\alpha - \left(r_\odot\over r_s\right)^\alpha
	\right)\right]
\ee
with $\rho_\odot = 0.4$ GeV/cm$^3$, $r_\odot = 8.3$ kpc, 
$\alpha=0.17$, $r_s=15$ kpc, consistent with best-fit values of 
$N$-body simulations,\cite{navarro}\  and a high value of the 
circular velocity $v_0=250$ km/s.   In this example, the heavy
contours show that there exist parameters leading to a large enough 
rate,
but these tend to disappear rapidly if one increases the values of 
$\delta M_{23}$ (since the $\chi_2$ states do not have enough kinetic
energy to produce $\chi_3$), or $\alpha$ or $r_s$ (since then $\rho$
becomes too small in the central region of the galaxy, reducing the
rate).  This can be compensated by decreasing $\delta M$ on the other
hand, as  illustrated in the right panel of fig.\ \ref{xdmres}.  
The shaded regions are ruled out by constraints on inverse
Compton gamma rays,\cite{papucci} as we discuss in the next section.
Fig.\ \ref{angle} shows that the more cuspy DM profile
with $\alpha=0.17$ gives a better fit to the angular distribution 
of the 511 keV signal.

\section{High energy $e^\pm$ from annihilations}

Although in refs.\ \cite{twist,nonabelian} we showed that the XDM mechanism
with inverted mass hierarchy can work for the 511 keV signal, we did
not consider whether it could also be compatible with the PAMELA and
Fermi lepton excesses.  The same model can also explain the high
energy leptons through annihilation to hidden sector gauge bosons,
$\chi_1\chi_1\to BB$, followed by the decays 
$B\to e^+e^-$\cite{meade}.  However, this scenario has come under increasing
pressure from various constraints, the most stringent being due to
inverse Compton scattering of $e^\pm$ on starlight in the galaxy,
which should produce $\gamma$ rays with energies up to several
hundred GeV.  Demanding that this new source not exceed recent
observations excludes the annihilating DM interpretation of Fermi
leptons unless the galactic DM density profile is less cuspy near the
center\cite{papucci} than is generally expected on the basis of $N$-body
simulations of halo evolution.\cite{navarro}\ \  This limit requires
taking small values of $\delta M\lsim 100$ keV in order to get 
a large enough rate for 511 keV $\gamma$ rays, as illustrated
in fig.\ \ref{xdmres}.

\begin{figure}[t]
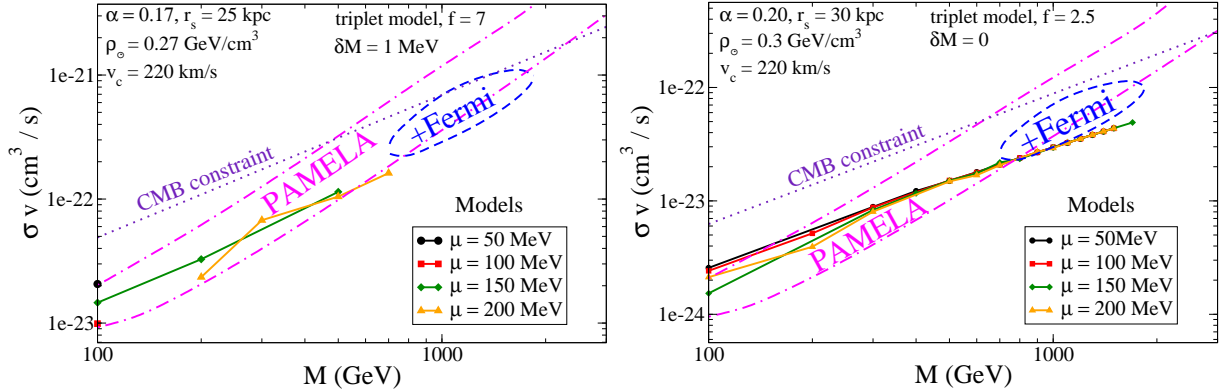

\centerline{\psfig{figure=fermi-a17-f7.eps,width=0.5\textwidth}
\hfil
\psfig{figure=fermi-a20-f25.eps,width=0.5\textwidth}} \caption{Allowed
regions for PAMELA and Fermi lepton excess in $\sigma v$-$M$ 
plane,$^{19}$\ 
and predictions of multistate DM annihilation that are compatible with
inverse Compton constraint.  Left: for Einasto parameters $\alpha =
0.17$, $r_s=25$ kpc, $\rho_\odot = 0.28$ GeV/cm$^3$; Right: 
for $\alpha = 0.20$, $r_s=30$ kpc, $\rho_\odot = 0.3$ GeV/cm$^3$.
$1/f$ is fraction of total DM mass density occupied by annihilating
DM ground state $\chi_1$.
}
 \label{fits}
\end{figure}
 
The ability of the models to explain the high-energy lepton
observations while respecting the IC constraints are summarized in
figure \ref{fits} taken from ref.\ \cite{CC}.  The left figure
is an example using a cuspy halo profile compatible with the PAMELA
and 511 keV excesses, at $M< 400$ GeV, while the right one shows the
result of a noncuspy profile where the PAMELA and Fermi excesses can
be marginally explained, but not the 511 keV.

\section{Conclusions}

We have found that annihilating multistate DM can explain two out of
three galactic cosmic ray anomalies, either PAMELA/Fermi or
PAMELA/INTEGRAL, but not all three simultaneously.   Although it is
possible to marginally predict all the correct rates using Einasto
profile parameter $\alpha=0.20$, the angular distribution of 511 keV
$\gamma$ rays is too wide in this case.   Of the two
possibilities, the PAMELA/INTEGRAL combination seems preferable from the standpoint
of the required DM halo parameters, since in this case we are able to
adopt standard values that are quite compatible with $N$-body
simulations of galactic structure evolution.  Moreover we can match
the anomalous lepton rates well for PAMELA/INTEGRAL.  The PAMELA/Fermi 
possibility requires stretching the halo parameters to their maximal
values, while only marginally giving a large enough rate of leptons,
yet a small enough rate of associated inverse Compton $\gamma$ rays.

\end{document}